\documentclass[aps,prb,twocolumn,showpacs,floatfix,nofootinbib]{revtex4}
\usepackage{amsfonts}
\usepackage{graphicx}
\usepackage{amsmath}
\usepackage{times}
\usepackage{amssymb}
\usepackage{color}
\usepackage[ colorlinks, bookmarks=false,citecolor=blue, linkcolor=red,urlcolor=magenta]{hyperref}

\newcommand{\diagram}[1]{\vcenter{\hbox{\includegraphics[scale=0.3]{./#1.pdf}}}}

\begin{document}
\title{Optimization schemes for unitary tensor-network circuit}
\author{Reza Haghshenas}
\affiliation{Division of Chemistry and Chemical Engineering, California Institute of Technology, Pasadena, California 91125, USA}
\date{\today}

\begin{abstract} 

An efficient representation of a quantum circuit is of great importance in achieving a quantum advantage on current Noisy Intermediate Scale Quantum (NISQ) devices and the classical simulation of quantum many-body systems. The quantum circuits are playing the key ingredient in the performance of variational quantum algorithms and quantum dynamics in problems of physics and chemistry. In this paper, we study the role of the network structure of a quantum circuit in its performance. We discuss the variational optimization of quantum circuit (a unitary tensor-network circuit) with different network structures. The ansatz is performed based on a generalization of well-developed multi-scale entanglement renormalization algorithm and also the conjugate-gradient method with an effective line search. We present the benchmarking calculations for different network structures by studying the Heisenberg model in a strongly disordered magnetic field and a tensor-network $QR$-decomposition. Our work can contribute to achieve the most out of NISQ hardware and to classically develop isometric tensor network states.  

\end{abstract}

\maketitle

\section{Introduction}
\label{Sec:introduction}

The field of tensor-network formalism has emerged as a novel toolbox to address the long-standing problems of modern condensed-matter physics.\cite{Verstraete:2008, Orus:2014} It offers an unbiased variational ansatz to simulate the physics of strongly entangled many-body systems, such as frustrated spins\cite{Corboz:2013, Wang:2016, Poilblanc:2017, Haghshenas:2018, Haghshenas:2018may, Niesen:2018,Haghshenas:2019May} and interacting fermions.\cite{Corboz:2010:April, Corboz:2014, zheng2017stripe} Tensor-network variational methods are not hampered by the exponential growth of the Hilbert space and the so-called sign problem (occurring in Quantum Monte Carlo algorithms), with the only essential limiting parameter being the amount of entanglement in the system. Since the low-energy states of physical systems obey entanglement area-law,\cite{Vidal:2003, Calabrese:2004, Plenio:2005} they could be efficiently simulated by a tensor-network ansatz in a polynomial time scale. However, tensor-network methods are not only limited to numerical simulation of many-body systems, but they are rapidly being extended to other areas of research such as the classification of novel phases of matter,\cite{Pollmann:2010, Chen:2011:Jan, Schuch:2011, Haghshenas:2014} machine learning and quantum computation,\cite{Stoudenmire:2016, Miles:2018, Feng:2019, McArdle:2019, Gray:2020, Motta:2020,  lin2020real} and variational (non-perturbative) approaches to quantum field theories.\cite{Verstraete:2010, Haegeman:2010, Haegeman:2013}

While mostly the tensor-network states are the underlying key ingredient in the variational ansatz; but in many physical systems, a unitary tensor-network circuit (uTNC) plays that key role. In a system that exhibits quantum many-body localization (MBL),\cite{Basko:2006, Fleishman:1980, Gornyi:2005}  violating the eigenstate thermalization hypothesis (ETH),\cite{Deutsch:1991, Srednicki:1994} the full energy spectrum could efficiently be represented by a uTNC as (almost) all eigenvectors obey an entanglement area law.~\cite{Hastings:2007:Area, Friesdorf:2015,Bauer:2013, Luitz:2015, Kjall:2014} One can efficiently approximate the unitary that diagonalizes an MBL Hamiltonian\cite{Pollmann:2016, Wahl:2017} by a local unitary circuit with finite depth. Additionally, in the context of quantum computation, a quantum algorithm is performed by a unitary quantum circuit including gates. One can think of this circuit as a uTNC and use well-developed tensor-network techniques such as efficient exact contractions, entanglement-truncation techniques, and optimization algorithms to systematically manipulate the circuit as desired.\cite{McArdle:2019, Motta:2020, lin2020real} Furthermore, in the context of projected entangled-pair states (PEPS) algorithms, a canonical form has been recently proposed based on tensor-network $QR$-decomposition. The tensor-network Q is made of a circuit of unitary and isometry, representing an isometric tensor-network circuit (iTNC).\cite{Haghshenas:2019, Zaletel:2020}

The optimization of uTNC is usually performed by minimizing a cost function by using conjugate-gradient-based methods.\cite{Alan:1998,Manton:2002} The computational cost of the ansatz scales linearly with the system size, whereby larger system sizes could be simulated. In the first proposal of the variational uTNC ansatz,\cite{Pollmann:2016} aimed at addressing MBL, the authors utilized a set of two-body unitary gates (rank-four tensors) arranged in a regular network to minimize the energy variance, which served as the cost function. It was shown that the accuracy of the ansatz rapidly improves with the depth of the network, denoted $\tau$. Recently, it has been discussed that a different scenario\cite{Wahl:2017} could significantly improve the accuracy: the main idea is to use $l$-body unitary gates in the same regular architecture as before, but the number of layers is fixed at $\tau=2$. To reduce the computational cost, the authors use a cost function based on the integrals of motion (instead of energy variance).

There remain many possible avenues to improve the variational uTNC ansatz and establish a standard method. The primary drawback of this approach is the poor convergence rates of conjugate-gradient algorithms, which significantly increases the computation cost. Specifically, for highly entangled excited states many iterations are required to obtain a converged result, limiting the control parameter $\tau$ and $l$. Additionally, it is unknown how different network structures play a role in the performance of the uTNC ansatz. One might utilize novel structures (aimed to more efficiently address entanglement) to improve the accuracy, without increasing the computational cost. Furthermore, a notable advantage of this method, compared to energy targeting methods,\cite{Khemani:2016, Yu:2017, Lim:2016,Serbyn:2016} is the ability to simulate the real-time dynamics of the system, which is of great interest.

In this article, we address the aforementioned possibilities by introducing new network structures and efficient optimization protocols, similar to that of multi-scale entanglement renormalization ansatz\cite{Vidal:2007} (MERA), to improve upon previous approaches. We use well-developed MERA algorithms~\cite{Evenbly:2009} to show the energy variance (cost function) can be locally evaluated efficiently. The energy variance is then minimized by two optimization protocols which significantly accelerate convergence rates: $(i)$ a linearizing algorithm used in MERA and $(ii)$ a conjugate-gradient algorithm with an efficient line-search method.\cite{Abrudan:2009} Both algorithms reduce computational cost as the system size and/or variational parameters increase. We further discuss that the MERA-like network structure not only takes into account larger correlations (resulting in better accuracy) but also could simply be used to study real-time dynamics.

We numerically benchmark the variational uTNC ansatz for the Heisenberg model with random magnetic fields~\cite{Pal:2010, Luitz:2015} and in a tensor-network $QR$-decomposition\cite{Haghshenas:2019} to check the validity of these methods. We compare the accuracy of different networks with that of previous ones and analyze the improvement in computational cost. A sample Python source code for the algorithms, presented in this paper, is available at \href{https://github.com/rezah/unitary-tensor-network-operator}{github.com/uTNC}.

\section{unitary tensor-network circuit ANSATZ}
\label{Sec:ANSATZ}
The main idea in tensor-network formalism is to efficiently represent a quantum state/operator in terms of local tensors (an object with several indices) connected by so-called virtual bonds. Tensor network representations allow us to manipulate the quantum state/operator through the individual tensors even for a large number of particles ($N \rightarrow \infty$). The tensors are connected through a specific network structure usually determined by intuition from the physical properties of the system. The network structure is designed to capture, at the least, the main global physics of the system such as scaling of entanglement entropy and correlation functions.\cite{Evenbly:2011} For instance, to faithfully describe a 1D gapless system, a 2D holographic network structure (produced by MERA) is required to generate correctly logarithmic entanglement scaling and algebraic fall-off of the correlation functions. On the other hand, 1D gapped systems respecting entanglement area law require a simple 1D tensor-network structure, produced by matrix product state (MPS). Importantly, the tensor-network structure plays a key role in the efficiency and accuracy of the resulting tensor-network ansatz.

\begin{figure}
  \begin{centering}
\includegraphics[width=1.0 \linewidth]{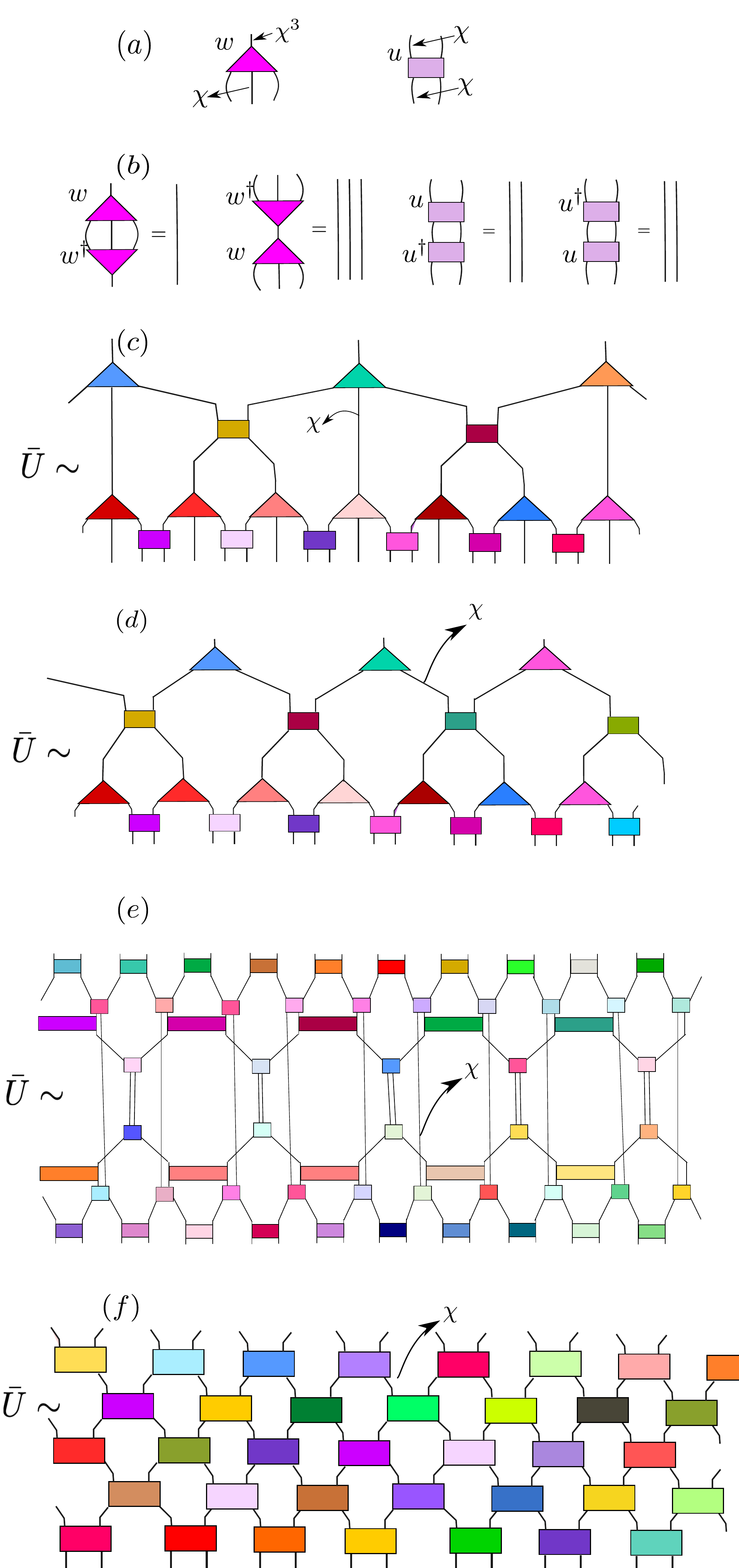}  \end{centering}
  \caption{(Color online) Tensor-network representation of many-body unitary $\bar{U}$ in terms of local unitary tensors $\{u, w\}$. $(a)$ Each triangular tensor $w$ is a $\chi^{3}\times \chi^{3}$ unitary matrix. The upper index have the bond dimension $\chi^{3}$, obtained from fusing three lower indices each with the bond dimension $\chi$. The square tensor $u$ is a $\chi^{2} \times \chi^{2}$ matrix, where each index takes the bond dimension $\chi$. $(b)$ The tensors $\{u, w\}$ are unitary, i.e. $u^{\dagger}u=uu^{\dagger}=\mathcal{I}$ and so on. $(c)$ A ternary uTNC $\bar{U}$ which approximately diagonalize a Hamiltonian. The bond dimension $\chi$, controlling accuracy of uTNC ansatz, is $8$. $(d, e, f)$ The binary, iregular and regular uTNC with bond dimensions $\chi=\{4, 2, 2\}$ and layer numbers (circuit depth) $\tau=\{4, 8, 5\}$, respectively. }
  \label{fig:MERAarchitecture}
\end{figure}

In this section, we aim to use a MERA-like network structure to build a unitary tensor-network circuit $\bar{U}_{\{u, w\}}$ to approximate the unitary $U$ that diagonalizes a Hamiltonian $H$
\begin{equation}
U \approx \bar{U}_{\{u, w\}}, \quad  U^{\dagger}HU=D,
\end{equation}
where $D$ is a diagonal matrix, whose diagonal elements are the eigenvalues of Hamiltonian $H$. The uTNC $\bar{U}_{\{u, w\}}$ is composed of local unitary tensors $\{u, w\}$
\begin{equation}
u^{\dagger}u=uu^{\dagger}=\mathcal{I}, \quad w^{\dagger}w=ww^{\dagger}=\mathcal{I},
\end{equation}
as shown in Fig.~\ref{fig:MERAarchitecture}(a, b). The tensors $\{u, w\}$ are then connected through a MERA network structure to construct $\bar{U}_{\{u, w\}}$, as depicted in Fig.~\ref{fig:MERAarchitecture}(c). Its explicit form is given by  
\begin{equation}
\bar{U}_{\{u, w\}}=\sum_{\tau, \bar{\tau}}  \mathcal{F}(\{u, w\}_{\tau, \bar{\tau}})   |\tau\rangle \langle \bar{\tau} |,
\end{equation}
where $|\tau\rangle=|\tau_{1}\cdots \tau_{N}\rangle$ forms a complete basis of $2^N$ states (the same for $|\bar{\tau}\rangle$) and $\mathcal{F}$ denotes tensor contraction. Since $\bar{U}_{\{u, w\}}$ approximately diagonalizes the Hamiltonian, then the eigenstates are given by $\bar{U}_{\{u, w\}} |\bar{\tau} \rangle$, i.e., 
\begin{equation}
|\psi_{\bar{\tau}}\rangle =\sum_{\tau}  \mathcal{F}(\{u, w\}_{\tau, \bar{\tau}})   |\tau\rangle,
\end{equation}
which reveals that all eigenstates $|\psi_{\bar{\tau}}\rangle$ are represented by a finite-range non-homogeneous MERA. In other words, $\bar{U}_{\{u, w\}}$ encodes all eigenstates into a finite-range non-homogeneous MERA. In this context, the tensors $\{u\}$ are the so-called unitary disentanglers removing short-range entanglement; while the tensors $\{ w \}$ transform a block of three spins into one single superspin. The tensors $\{u, w\}$ define a coarse-graining transformation which maps the spins at the $(\tau)$th layer, with Hilbert space dimension $\chi_{\tau}$, into superspins at the $(\tau+1)$th layer, with a larger Hilbert space dimension $\chi_{(\tau+1)}=\chi^3_{(\tau)}$. In order to avoid this growth of Hilbert space, the number of layers should be limited to finite values, hence the finite-range MERA. 

All tensor-network variants of MERA, such as the so-called binary, modified binary, ternary, and tree tensor-network structures could be used as a network in the uTNC~\cite{Evenbly:2013}---obtained by replacing the isometric tensors by unitary ones. We have shown some network structures in Fig.~\ref{fig:MERAarchitecture}(c-f) used in this paper as a uTNC ansatz. The irregular uTNC, made of two connected binary MERA (one has been rotated) through long bonds, has an important feature that allows the light cone to grow exponentially with the uTNC's depth $\tau$, compared to regular uTNCs, where the light cone grows linearly. Thus it is expected that the irregular network structure could better capture long-range properties. 

The proposed uTNCs provide varying levels of accuracy and computational cost so that the most efficient ansatz can be empirically chosen. Furthermore, the MERA optimization methods can be explicitly generalized to uTNC ansatz, as we describe it in Sec.~\ref{Sec:OPTIMIZATION}. For the sake of simplicity, we use the ternary uTNC, depicted in Fig.~\ref{fig:MERAarchitecture}(c), to discuss the algorithms and evaluation of the real-time dynamics.

\section{OPTIMIZATION ALGORITHMS}
\label{Sec:OPTIMIZATION}
\begin{figure}
  \begin{centering}
\includegraphics[width=1.0 \linewidth]{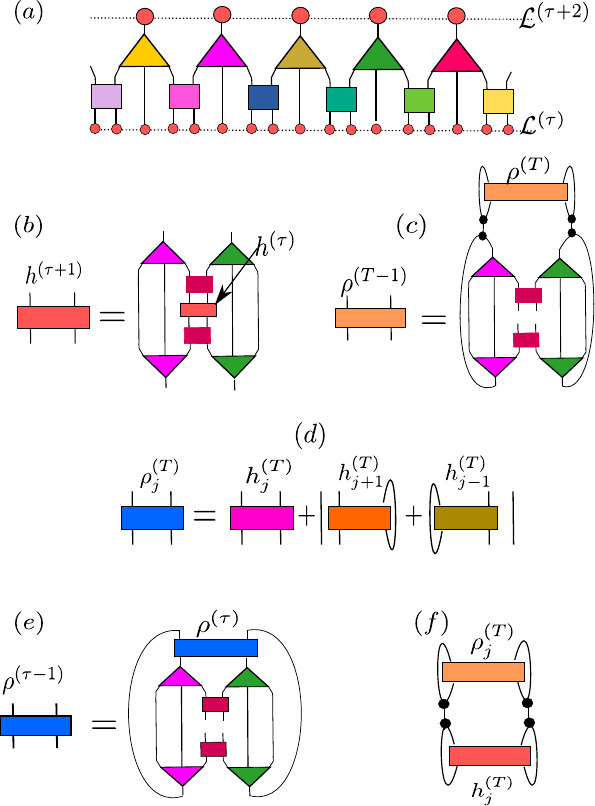}  \end{centering}
  \caption{(Color online) $(a)$ a two-layer transformation, composed of the unitary tensors $\{w, u\}$, maps three spins (small red circles) of lattice $\mathcal{L}^{\tau}$ into a single superspin (big red circles) of lattice $\mathcal{L}^{\tau+2}$. Tensor-network representation of $(b)$ ascending superoperator $\mathcal{A}$ for layer $\tau$ and $(c)$ descending superoperator $\mathcal{D}$ for the topmost layer $T$. The black circles represent delta function $\delta_{ijm}$. $(d)$ $\rho^{(T)}$ is obtained from the transformed operators $h^{(T)}$. $(e)$ The descending superoperator $\mathcal{D}$ at the layer $\tau<T$. $(f)$ The tensor contraction $\mathcal{F}(h^{(T)} \rho^{(T)})$ has a simple graphical representation.} 
  \label{fig:Ascending}
\end{figure}

We discuss some `improved' optimization methods to approximate the unitary $ U \approx \bar{U}_{\{u, w\}}$ based on mimizing a cost fucntion. As we aim to obtain the full spectrum of a particular Hamiltonian $H$, we consider the energy variance $\sigma^{2}$ as cost function \cite{Pollmann:2016}  and seek to minimize it with respect to the tensors $\{u, w\}$. The cost function $\sigma^{2}$ reads
\begin{equation}
 \sigma^{2}=\mathrm{Tr}(H^{2})-\sum_{i=1}^{2^{N}} [(\bar{U}^{\dagger}H\bar{U})^{2}]_{(i,i)},
\end{equation}
where symbol $(i, i)$ stands for diagonal elements of a matrix. The cost function $\sigma^{2}$ is the energy variance summed over all approximate eigenstates. Note $\sigma^{2}=0$ strictly indicates that $\bar{U}_{\{u, w\}}$ exactly represents all true eigenstates. The first term does not play any role in optimization procedure, so we ignore it and seek to maximize the second term, which can be efficiently calculated (computational time scales linearly with system size $N$). This is done by the so-called $\emph{ascending}$ and $\emph{descending}$ superoperators (similar to those appearing in standard MERA algorithms) that are used to move local operators up and down to different layers, as illustrated in Fig.~\ref{fig:Ascending}(a) and detailed in the following section.

\subsection{Efficient representation of cost function}
\label{Sec:Ascending}
We assume that the Hamiltonian $H$ only includes short-range interactions, i.e. $H=\sum_{i} h_{i,i+1}$. The local terms $h$ could be mapped to upper layers $\tau$ by applying $\emph{ascending}$ superoperator $\mathcal{A}$ as shown in Fig.~\ref{fig:Ascending}(b). This produce a sequence of operators, each defined on different layers
\begin{equation}
h^{(0)}  \stackrel{\mathcal{A}}{\rightarrow}  h^{(1)} \stackrel{\mathcal{A}}{\rightarrow} ~\cdots~ \stackrel{\mathcal{A}}{\rightarrow} h^{(T)},
\end{equation}
where upper indices specify the layer $\tau$ ($0 \leq \tau \leq T$). The term $h^{(0)}$ shows local original interaction $h^0 \equiv h$, while $h^{(\tau)}$ represents a two-body interaction defined on two adjacent superspins with local Hilbert-space dimension $\chi_{\tau}$. By using the superoperator $\mathcal{A}$, we simply find that $\bar{U}^{\dagger}H\bar{U}=\sum^{\frac{N}{3^T}}_{j=1} h_{j}^{(T)}$. The square of this quantity, which appears in the cost function takes a simple form, given by
\begin{equation}
\sum_{i=1}^{2^{N}} ( \sum_j h_{j}^{(T)}  \sum_m h_{m}^{(T)} )_{(ii)}=\chi_{T}^{(\frac{N}{3^T}-2)}\sum_{j=1}^{\frac{N}{3^T}} \mathcal{F}(h_{j}^{(T)} \rho_{j}^{(T)} ),
\end{equation}
where $\rho_{j}^{(T)}$ is obtained from $\{ h_{j-1}^{(T)},h_{j-1}^{(T)}, h_{j+1}^{(T)}\}$ as shown in Fig.~\ref{fig:Ascending}(d). The tensor contraction $\mathcal{F}$ is defined in Fig.~\ref{fig:Ascending}(e) which is slightly different from the matrix trace. The main reason for such simplification is due to the unitary constraint (annihilating most tensors to the identity) which makes the computational time linear in the system size, i.e. $\sum_{i=1}^{2^{N}}  \equiv \chi_{T}^{(\frac{N}{3^T}-2)} \sum_{i=1}^{\chi_{T}^{2}} $. Therefore, the final form of the cost function $\sigma^{2}$ is given by
\begin{equation}
\sigma^{2}=const-\chi_{T}^{(\frac{N}{3^T}-2)} \sum_{j=1}^{\frac{N}{3^T}}\mathcal{F}(h_{j}^{(T)} \rho_{j}^{(T)} ).
\end{equation}
In this form, only operators at the topmost level $T$ appear, but in the optimization procedure, we need a similar expression for each layer $\tau$. To do that, we need to define the counterpart of $\emph{ascending}$ superoperators $\mathcal{A}$ to move the operators to lower layers, i.e., $\emph{descending}$ superoperators $\mathcal{D}$ as depicted in Fig.~\ref{fig:Ascending}(c). Similarly, we obtain a sequence of operators in different layers
\begin{equation}
\rho^{(0)}  \stackrel{\mathcal{D}}{\leftarrow}  \rho^{(1)} \stackrel{\mathcal{D}}{\leftarrow} ~\cdots~ \stackrel{\mathcal{D}}{\leftarrow} \rho^{(T)}.
\end{equation}
By using both superoperators $\mathcal{A}$ and $\mathcal{D}$, we can finally express the cost function in layers $\tau< T$
\begin{equation}
\sigma^{2}=const-\chi_{T}^{(\frac{N}{3^T}-2)}\sum_{j=1}^{\frac{N}{3^\tau}} \mathrm{tr}(h_{j}^{(\tau)} \rho_{j}^{(\tau)} ),
\end{equation}
where the symbol `$\mathrm{tr}$' stands for the matrix trace. The leading computational cost (in all steps) scales as $\mathcal{O}(\chi^9_{T-1})$ for the ternary uTNC and as $\mathcal{O}(\chi^{10}_{T-1})$ for the binary uTNC.

\subsection{Conjugate-gradient method with efficient line-search algorithm}
\label{Sec:CGmethod}

With an efficient method for computing the cost function, we now need to minimize it by sequentially optimizing the local unitary tensors. An iterative strategy provides an efficient way to do that: at each step, one tensor is optimized while others are held fixed, repeating this for all tensors until the cost function does not change significantly. However, the non-linear nature of the cost function and the unitary constraint for each tensor makes this optimization procedure challenging. Well-developed algorithms exist to handle this optimization, specifically variants of the gradient method, such as steepest-descent (SD), conjugate-gradient (CG), and quasi-Newton algorithms. 

The basic idea in the SD method is to minimize the cost function in the direction of the gradient. In each iteration $i$, the new unitary tensor $u_{i+1}$ is obtained by 
\begin{equation}
u_{i+1}=e^{-\alpha g_{i}} u_{i},
\end{equation} 
where $\alpha$ is step-size parameter and $g_i$ is the Riemannian gradient direction (an skew-symmetric matrix)
\begin{equation}
g_i=u_{i}Y^{\dagger}_{u_{i}}-Y_{u_{i}}u^{\dagger}_{i}, \quad Y_{u_{i}}=\partial_{u_{i}}\sigma^{2},
\end{equation}
where tensor $Y_{u_{i}}$ is the so-called $\emph{environment}$ tensor, obtained by contracting all tensors around $u_{i}$. In our case, it has a simple tensor-network representation, thanks to simple form of the cost function obtained in Sec.~\ref{Sec:Ascending}, which is calculated once in each step $i$. After obtaining the gradient direction, the next step is to find the optimal value of $\alpha$, referring to the line-search algorithm parameter. The standard approach is to minimize the cost function $\sigma^{2}(\alpha)$ by using the above equation while systematically decreasing $\alpha$, often according to certain criteria, such as the Armijo condition. Unfortunately, in practice, finding optimal $\alpha$ requires multiple evaluations of cost function, making the line-search algorithm the main computational bottleneck of both the SD and CG methods.

We could improve the gradient optimization implementation by using a better choice for the search direction. In the SD method, the minimum point is usually approached in a zig-zag route (takes ninety-degree turns at every iteration), but in the CG method, the minimum point is often reached via a relatively direct path. There, the new search direction is determined by a proper choice between current and previous search directions, thus the new unitary tensor $u_{i+1}$ is obtained by    
\begin{equation}
u_{i+1}=u_{i}-\alpha\bar{g}_{i},
\end{equation}
where $\bar{g}_{i}=g_{i}-\beta_{i} g_{i-1}$. The parameter $\beta_{i}$ is determined by Polak-Ribi\'{e}re formula (or other prescriptions), i.e. $\beta_{i}=\frac{<g_{i}-g_{i-1},g_{i}>}{<g_{i-1}>}$, where the notation $<>$ denotes the matrix norm. In principle, the CG algorithm requires an accurate search direction, thus a line-search algorithm plays a crucial role in the efficiency/accuracy.

A strategy to make the line-search method more efficient, avoiding multiple evaluations of the cost function, is to directly approximate the first-order derivative of the cost function and solving equation $\partial_{\alpha} \sigma^{2}(\alpha)=0$. The zero points of this equation are the optimal step-size $\alpha$. We use a low-order polynomial approximation (up to $p$th order) and take the corresponding smallest positive root as step-size value, which significantly reduces computational time by avoiding cost function evaluations.\cite{Traian:2009} This method provides accurate results while the cost function (and its derivatives) would be almost periodic with respect to $\alpha$. The function $\sigma^{2}(\alpha)$ is called $\epsilon$-almost periodic, if there exists a real number T, so that $|\sigma^{2}(\alpha+T)-\sigma^{2}(\alpha)|<\epsilon, \forall \alpha$. The steps to find an optimal step size are the following (for more detail see Ref.~\onlinecite{Traian:2009}):

\begin{itemize}

    \item[(a)] \underline{\it Period of the cost function:}  Compute the largest eigenvalue of the $g_i$, i.e. $|w_{max}|$. The largest polynomial degree of $u_i$, appearing in the cost function, determines the order of the cost function, denoted $q$. The parameter $q$ is equal to $4$ and $2$ in the energy-variance cost function and the tensor-network $QR$-decomposition, respectively. The period of cost function is then obtained by $\frac{1}{T}=\frac{q|w_{max}|}{2\pi}$,
   
    \item[(b)] \underline{\it  Sampling cost function by equi-spaced points:} Calculate matrices $R_k=e^{-\mu_k g_i}$, where $\mu_k \in \{ 0, \frac{T}{p},\frac{2T}{p}, \cdots, T  \}$ and $p$ is the low-order polynomial approximation parameter controlling accuracy. For $k \in \{0, \cdots, p\}$, we obtain the values of first-order derivative of cost functions  $\eta_{k}= -2  tr(\partial_{u_i}\sigma^{2}(R_ku_i)   u^{\dagger}_iR^{\dagger}_kg^{\dagger}_i)$ for $p$ equi-spaced points,
    
    \item[(c)] \underline{\it  Polynomial Coefficients:} Obtain the $p\times p$ matrix $z_{mn}=\mu^{n}_{m}$, and $p\times 1$ matrix $b_m=\eta_{m}-\eta_{0}$, where indices $m,n \in \{1,\cdots, p\}$, to compute coefficients $a=z^{-1}b$,

    \item[(d)] \underline{\it  Step Size:} By finding the smallest real positive root $\rho_{min}$ of  $a_0+a_1 x+\cdots+a_{p}x^{p}=0$, where $a_0=\eta_{0}$ and $a_1,\cdots, a_p$ is obtained from previous step $(c)$, we can estimate the step size.
   
\end{itemize}

If there is no solution to the polynomial equation, we need to increase $T$ or use alternative line-search algorithms to find step-size $\alpha$. In practice, we find that this algorithm works quite well (even with small values of the polynomial order $p \sim 3-4$) with the type of cost functions we are dealing with. The python code for this implementation is presented at \href{https://github.com/rezah/unitary-tensor-network-operator/blob/master/optimize.py}{github.com/uTNC/CG-Poly}.

\begin{figure}
  \begin{centering}
\includegraphics[width=0.90 \linewidth]{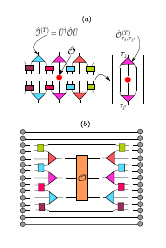}  \end{centering}
  \caption{(Color online) The real-time dynamics by using one-layer ternary uTNC. $(a)$ The term $\bar{U}^{\dagger} \hat{\mathcal{O}} \bar{U}$ finds a simple form due to annihilation of unitary operators to identity. $(b)$ The time evolution of a local operator $\langle \psi(t) | \hat{\mathcal{O}} |\psi(t)\rangle$ could be evaluated at the time $t$ by contracting this tensor-network diagram with computations cost $\mathcal{O}(\chi_{T-1}^{9})$. We represent the initial state with a MPS. The black circles represent delta function $\delta_{ijm}$. } 
  \label{fig:timeoperator}
\end{figure}

\subsection{Linearizing algorithm}
\label{Sec:Linearizing}

We describe an alternative method based on a linearizing application to optimize the cost function. The difficulty in minimizing the cost function mostly lies in its non-linear dependence upon the unitary tensors and the unitary constraint. The cost function $\sigma^{2}$ is a quartic function with respect to $u_{i}$, including, at the most, fourth-degree polynomial terms. The basic idea to simplify the optimization procedure is to temporarily make the cost function `linear' with respect to $u_{i}$, holding fixed all other tensors. To do that, we follow a strategy similar to one adopted in MERA simulation. We rewrite the cost function as follows
\begin{equation}
\sigma^{2}  \sim  \operatorname{tr}(Y_{u_{i},u^{\dagger}_{i}}u_{i}),
\end{equation}
where $Y_{u_{i},u^{\dagger}_{i}}$ is again the $\emph{environment}$ tensor. Note $Y_{u_{i},u^{\dagger}_{i}}$ strictly depends on $\{u_{i}, u^{\dagger}_{i}\}$, but the basic idea is to temporary assume $Y_{u_{i},u^{\dagger}_{i}}$ to be independent of them and accordingly minimize the cost function $\sigma^{2}$. The exact solution of this minimization problem is given by ${u}_{i+1}=-v^{\dagger}w$, where $w$ and $v$ are determined by singular value decomposition of the environment tensor $Y_i=w^{\dagger}sv$. We then repeat this process until ${u}_{i}$ converges. The steps for the linearizing algorithm are as follow:
\begin{itemize}
    \item[(a)] \underline{\it  $\emph{Environment}$ tensor:}  Compute the $\emph{environment}$ tensor of $u_i$ by contracting all tensors in $\sigma^2$, excluding $u_i$, then perform singular value decomposition to split it into $Y_{u_{i},u^{\dagger}_{i}}=w^{\dagger}sv$.  
    \item[(b)] \underline{\it  Update:} Choose $u_{i+1}=-v^{\dagger}w$ and replace $u_{i+1} \rightarrow u_{i}$  
    \item[(c)] \underline{\it  Evaluation:} Evaluate cost function $\sigma^{2}$ and return to step-(a) if it does not meet convergence criteria. 
\end{itemize}
The computational bottleneck of the linearizing method is computing the $\emph{environment}$ tensor. Since it does not require a line-search algorithm, this increases significantly the convergence rate. However, it is not generally guaranteed to provide accurate results, hence, its accuracy/validity should be empirically examined---as we observe in some cases it fails to find global minima, getting stuck in local minima, see Sec.\ref{Sec:nr}.

\section{TIME-EVALUATION ALGORITHM}
\label{Sec:timeevaluation}
In this section, we study the long-time dynamics of a local quantity by using ternary uTNC. We expect $\bar{U}$ (after the optimization procedure) to represent accurately $U$, hence
\begin{equation}
e^{-i t H}  \approx   \bar{U}^{\dagger}   e^{-i t D}   \bar{U} , 
\end{equation}
Our goal is to study the following equation
\begin{equation}
\langle \psi(t) | \hat{\mathcal{O}} |\psi(t)\rangle \approx \langle \psi(0)| \bar{U} e^{iDt} \bar{U}^{\dagger} \hat{\mathcal{O}} \bar{U} e^{-iDt} \bar{U}^{\dagger}|\psi(0)\rangle,
\end{equation}
where $\hat{\mathcal{O}}$ stand for a local operator, e.g., defined on one site and $|\psi(0)\rangle$ represents the initial quantum state at time $t=0$. The diagonal matrix $D$ requires calculating $2^N$ eigenvalues, which grows exponentially, thus further simplification is still required. By using the $\emph{ascending}$ superoperator $\mathcal{A}$, as shown in Fig.~\ref{fig:timeoperator}(a), we obtain
\begin{equation}
\hat{\mathcal{O}}^{(T)}=\bar{U}^{\dagger} \hat{\mathcal{O}} \bar{U},
\end{equation}
where $\hat{\mathcal{O}}^{(T)}$ is still a local operator defined on one superspin at the last layer $T$. In addition, we could find that $e^{iDt}  \hat{\mathcal{O}}^{(T)} e^{-iDt}$ has a simple form defined on three adjacent superspins, given by
\begin{align*}
\bar{\mathcal{O}}^{(T)}&_{ \tau_{j'-1}, \tau_{j'}, \tau_{j'+1}; \tau_{j-1}, \tau_{j}, \tau_{j+1}}= \\ 
&e^{-it(h^{(T)}_{\tau_{j-1}, \tau_j }+h^{(T)}_{\tau_j, \tau_{j+1}}-h^{(T)}_{\tau_{j'-1}, \tau_{j'}}-h^{(T)}_{\tau_{j'}, \tau_{j'+1}})}       
\hat{\mathcal{O}}^{(T)}_{\tau_j, \tau_{j'}}  . 
\end{align*}
Note that $\{\tau_{j-1}, \tau_{j}, \tau_{j+1}\}$ represent the Hilbert space of three adjacent superspins with dimension $\chi_T$. The key point for such a simplification is that $D$ is expressed in terms of the local operators, i.e., $D_{\tau_1 \cdots \tau_N}=\sum_{i} h_{\tau_i, \tau_{i+1}}^{(T)} $,
\begin{equation} 
\label{EQ:SplittingTR}
\diagram{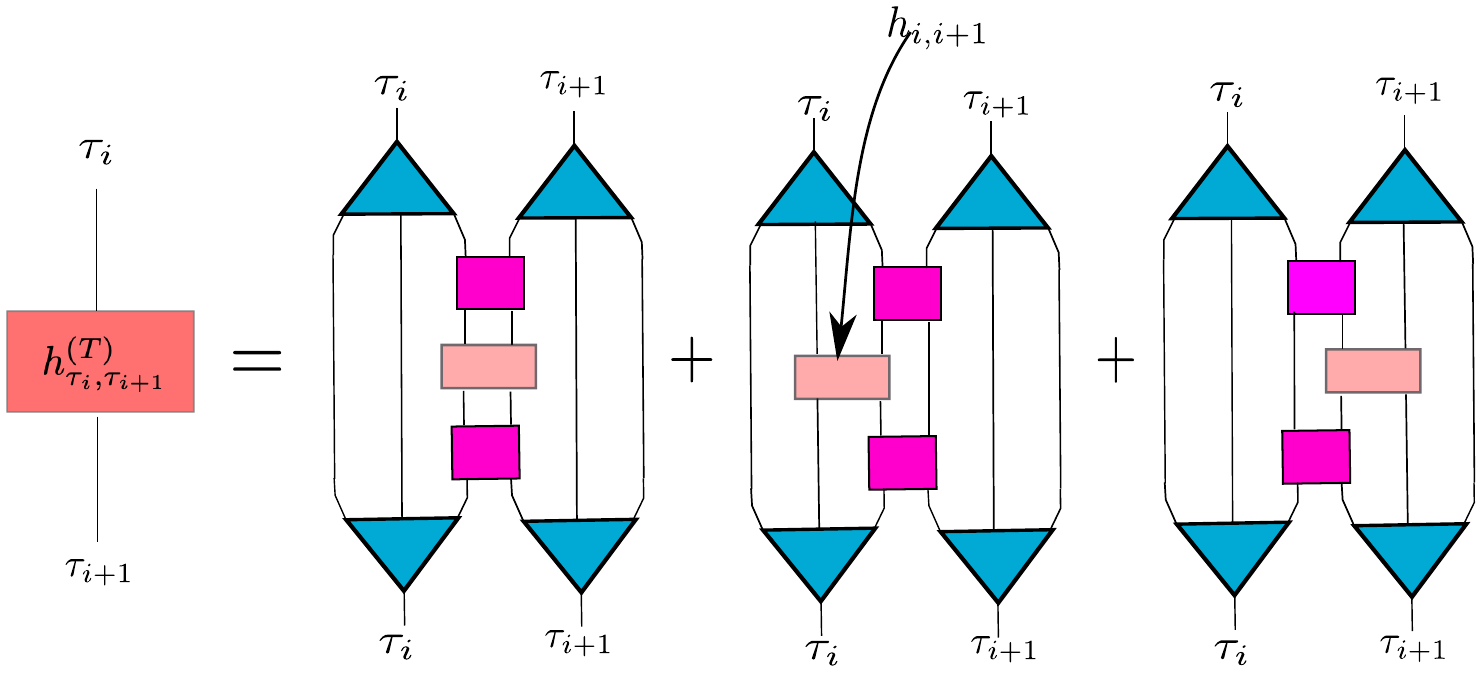},
\end{equation}
therefore, the time evolution of the local operator has the form
\begin{equation}
\langle \psi(t) | \hat{\mathcal{O}} |\psi(t)\rangle=\langle \psi(0)|U \bar{\mathcal{O}}^{(T)} U^{\dagger}|\psi(0)\rangle,
\end{equation}
with a simple tensor-network representation, shown in Fig.~\ref{fig:timeoperator}(b), where most of the unitary tensors are annihilated to identity. Similarly, one can find an efficient tensor-network representation for time-evaluation of the wave function $e^{-i t H} |\psi(0)\rangle$.

\section{NUMERICAL RESULTS}
\label{Sec:nr}

We analyze the algorithms presented by studying the Heisenberg chain with random magnetic fields given by the Hamiltonian
\begin{equation}
H= \sum_{i}\overrightarrow{S}_{i}\cdot\overrightarrow{S}_{i+1}-h_{i}S^{z}_{i},
\end{equation}
where $\overrightarrow{S}$ are spin-$1/2$ operators. The fields $h_{i}$ are drawn from a uniform distribution $[-W, W]$, where $W$ is called the disorder strength inducing different many-body phases. Exact diagonalization studies\cite{Pal:2010, Luitz:2015} ($N  \leq 20$) have predicted an ETH phase for $W \leq 3.5$ and a MBL phase for $W \geq 3.5$.

Additionally, we test the iTNC ansatz in a tensor-network $QR$-decomposition of a PEPS column $M$, see Fig.~\ref{fig:Conv}(a, b). The tensors in Q are represented by a circuit, which is made of a few layers of unitary gates with a layer of isometric tensors on the top. We can use the network structures and optimization schemes introduced in this paper to benchmark the accuracy of tensor-network $QR$-decomposition and study possible improvements. Our initial PEPS is constructed from the PEPS ground state of the two-dimensional Heisenberg model on a $l_y\times l_y$ square lattice.\cite{Lubasch:2014} The PEPS bond dimension is denoted by $\bar{D}$.

\begin{figure}
  \begin{centering}
\includegraphics[width=1.0 \linewidth]{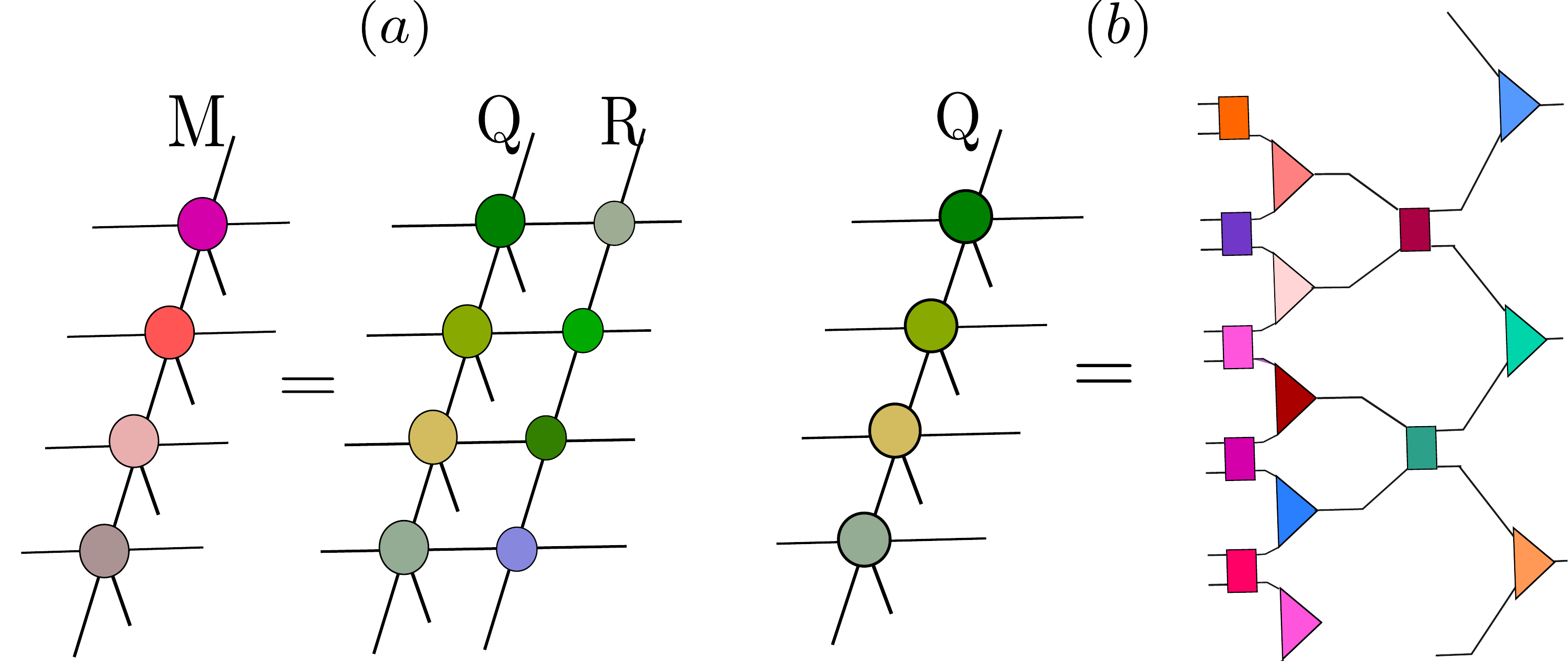}  \end{centering}
  \caption{(Color online) (a) A decomposition of a bulk PEPS column $M \approx QR$. $(b)$ The tensor-network $Q$ is parameterized by iTNC with a binary network structure.} 
  \label{fig:Conv}
\end{figure}

\subsection{Comparing the convergence rate of the optimization methods}
In this section, we benchmark the optimization methods introduced in Sec.~\ref{Sec:CGmethod} and Sec.~\ref{Sec:Linearizing}. We begin by studying energy variance $\sigma^{2}$ as a function of iteration number. The results are for the Heisenberg model with disorder strength $W=6$, located in the MBL phase. Note that the optimization methods could be generally applied to different types of geometrical networks and cost functions. 

\begin{figure}
  \begin{centering}
\includegraphics[width=1.0 \linewidth]{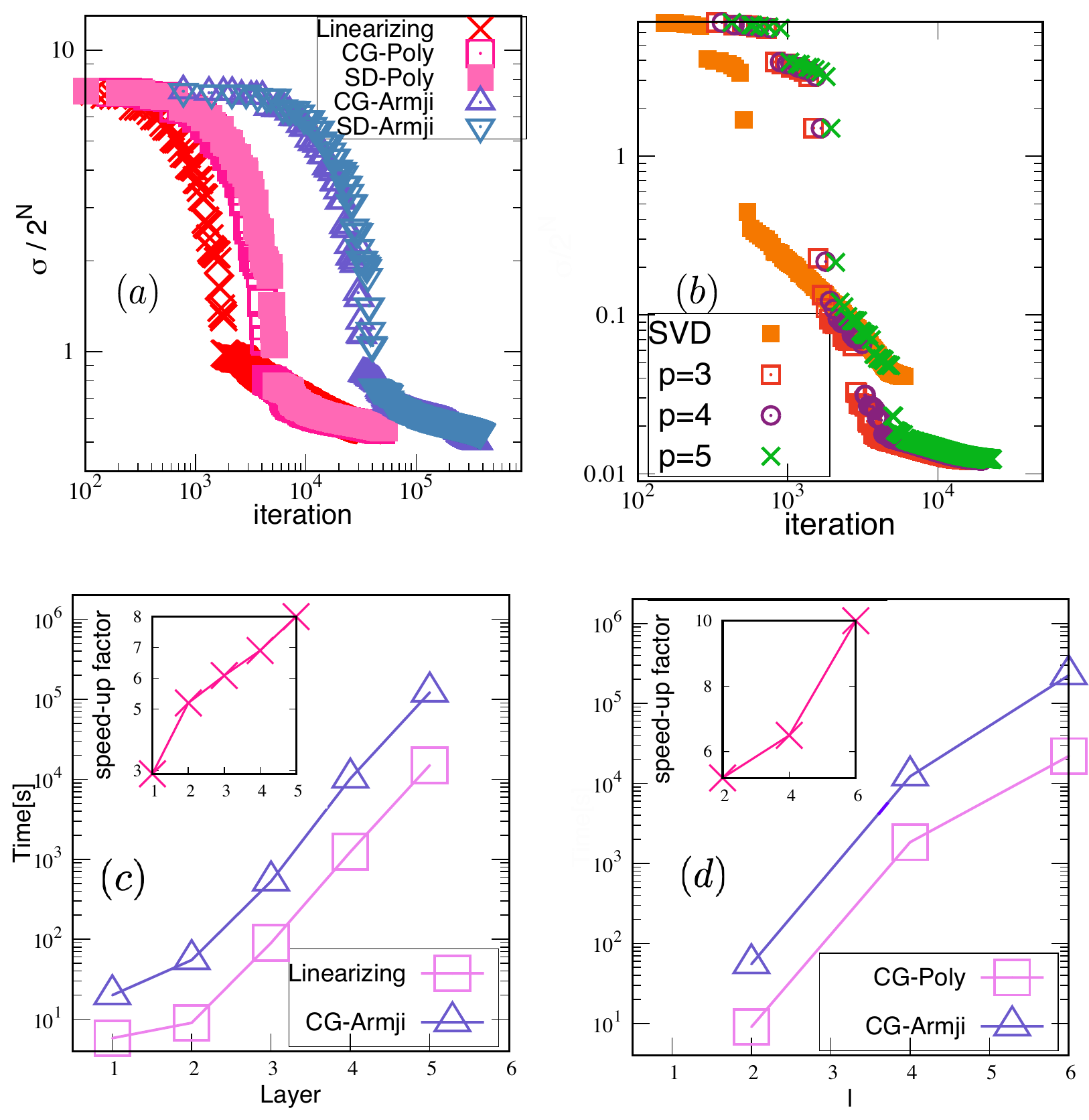}  \end{centering}
  \caption{(Color online) A comparison between the optimization methods of CG and SD with different line-search algorithms and the linearizing algorithm. $(a)$ Log-log plot of the cost function $\sigma^{2}/2^{N}$ versus iteration number for system size $N=32$ and disorder strength $W=6$. $(b)$ The same plot with different values of polynomial order $p$ used in CG method. The symbol `SVD' stands for singular value decomposition used in the linearizing algorithm. The system size and disorder strength are $N=20, W=6$. $(c)$ The averaged running time (seconds) as a function of circuit depth $\tau$ for the regular uTNC. The inset show the computational speed-up of linearzing method compared to CG with Armijo search-line algorithm. $(d)$ The averaged running time as a function of block size $l$ (circuit depth is fixed at $\tau=2$) for $N=32$. The inset shows the computational speed-up of CG method with polynomial line-search algorithm versus Armijo algorithm.} 
  \label{fig:Cost-rate}
\end{figure}

We apply the optimization methods to a regular architecture, see Fig.~\ref{fig:MERAarchitecture}(f), with circuit depth $\tau=5$. We first initialize the two-body unitary tensor by identity $\{ u \}=\{ \mathcal{I}\}$, then start the optimization sequentially by sweeping through the local tensors. We plot the results in Fig.~\ref{fig:Cost-rate}(a), where we observe that the linearizing algorithm and CG with an effective line-search provide accurate results, similar to CG with Armijo linea-search. Both algorithms approach the same accuracy level similar to that of the CG/SD with Armijo line-search, but with much fewer iterations. In this case (the type of our cost function), we do not observe a significant difference between CG and SD with the same line-search algorithm. In Fig.~\ref{fig:Cost-rate}(b), we seek how the CG method performs with different values of polynomial order $p$ for a binary uTNC. It shows that a small polynomial order $p=3$ is enough to efficiently reach converged results. In this case, we observe that the linearizing method gets stuck in a local minimum. Empirically, we find that the linearizing method might not be able to find converged results for the ternary and binary uTNCs (where the number of variational parameters per tensor increases rapidly by $\tau$), but it is accurate for the regular and irregular uTNCs. The CG with effective line-search seems to converge to the actual minimum always. This is an important advantage of the CG with effective line-search compared to other algorithms, providing huge speed-up without sacrificing accuracy.

To estimate the computational speed-up of the improved optimization schemes, we plot actual running time (directly calculated on a PC with $6$ core processor) $t$ averaged on 50 realizations versus the number of layers $\tau$. Empirically, we notice that the linearizing algorithm is faster than CG with effective line-search, and both algorithms are much faster than CG with an Armijo line-search, see Fig.~\ref{fig:Cost-rate}(c) and Fig.~\ref{fig:Cost-rate}(d). We see that this behavior remains the same for all structures, as convergence-rate/accuracy is almost independent of tensor-network architecture. As expected, by increasing the depth, we obtain a larger speed-up factor, up to $\sim 8, 10$. We expect to obtain larger factors by increasing system size $N$ and $\tau$ compared to the CG/SD with Armijo line-search. In our calculation, we have set polynomial order $p \sim 3, 4$, as we previously noted that a small polynomial order is enough to obtain converged results in most cases. In challenging cases, we might need to increase it up to $p \sim 4, 5$ to ensure the results are converged.

\subsection{The accuracy of the tensor-network architectures}
We study the performance of the tensor-network architectures by studying the energy variance $\sigma^2$. In Fig.~\ref{fig:Conv}(a), the variational power of the different uTNCs (represented in Fig.\ref{fig:MERAarchitecture}(c-f)) has been shown for a single realization for $N=20$ and $W=8$. We plot the cost function versus iteration number for the same initial guess for all architectures. It is seen that the binary and ternary uTNC outperform the regular and irregular ones. The main reasons for better performance in the binary and ternary structures are (i) having a larger number of variational parameters (per tensor) and (ii) effective connectivity of tensors efficiently capture the localization length of the system.\cite{Evenbly:2011, Evenbly:2013, Wahl:2017}
\begin{figure}
  \begin{centering}
\includegraphics[width=1.0 \linewidth]{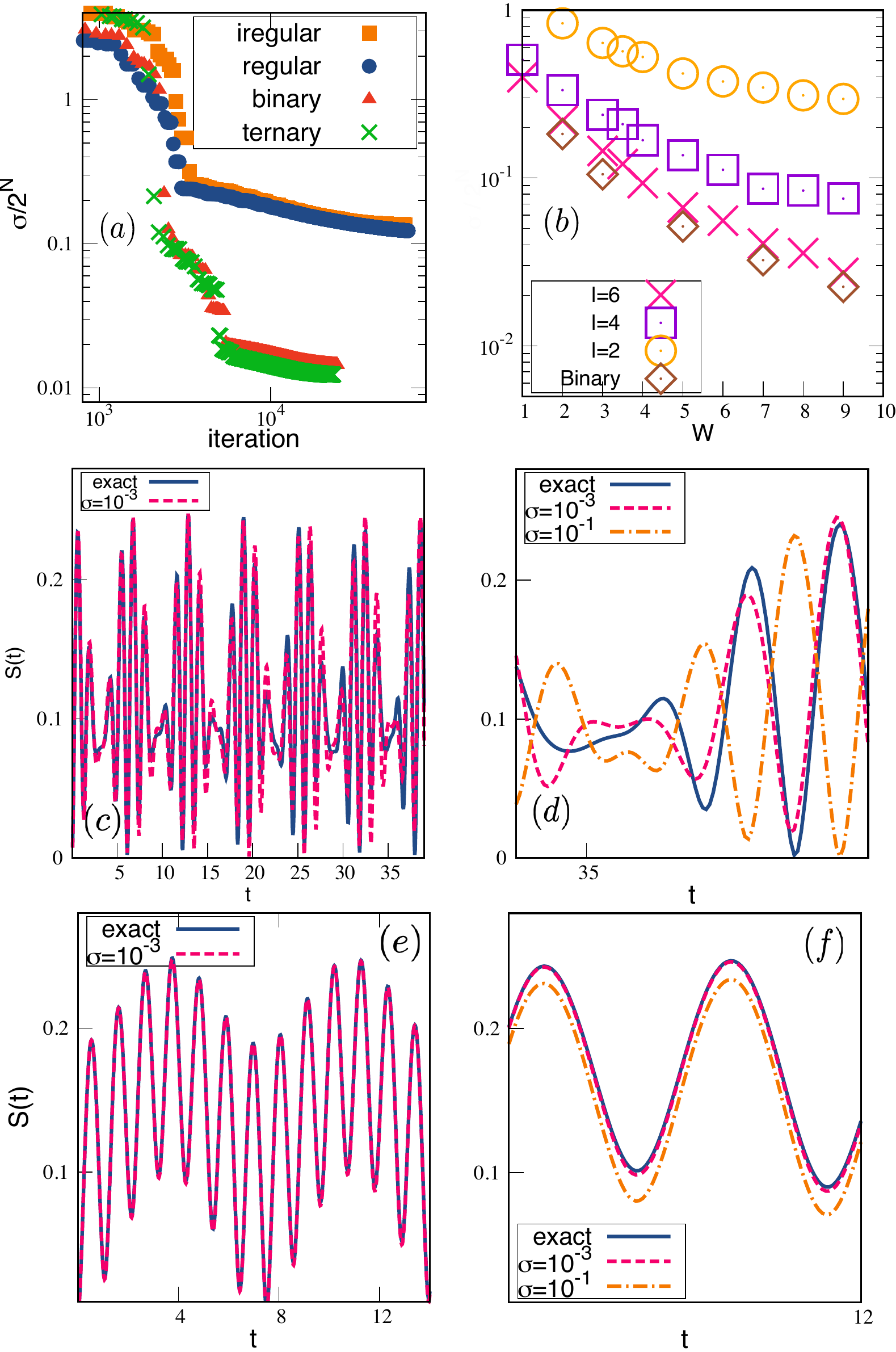}  \end{centering}
  \caption{(Color online)  $(a)$ Log-log plot of the cost function $\sigma^{2}/2^{N}$ versus iteration number for the networks shown in Fig.\ref{fig:MERAarchitecture}. The system size and disorder strength are set to $N=20, W=8$. $(b)$ The cost function $\sigma^{2}/2^{N}$ averaged over $40$ realizations as a function of $W$ (with $N=32$) for different architecture of the uTNC ansatz $\bar{U}_{\{u\}}$. The leading computaional cost of the binary uTNC ansatz is the same as (smaller) the ansatz in Ref.~\onlinecite{Wahl:2017} with $l=4$ ($l=6$). We observer the the binary uTNC ansatz provides better accuracy in all range of $W$. $(c, d)$ The time dependent entanglement entropy $S(t)$ verses time for $N=\{10\}$ and $W=8$, respectively in different time windows. $(e, f)$ The same results for $N=\{18\}$}. 
  \label{fig:Conv}
\end{figure}

Further, the binary uTNC could outperform the ansatz introduced in Ref.~\onlinecite{Wahl:2017}. To this end, we present results of the cost function versus the disorder strength for the regular uTNC with two layers $\tau=2$ and varying block sizes $l$ (exactly similar to the one used in Ref.~\onlinecite{Wahl:2017}) and a binary uTNC with $\chi=4$ shown in Fig.~\ref{fig:MERAarchitecture}(d). The results are presented in Fig.~\ref{fig:Conv}(b), where we observe that the binary ansatz provides better accuracy compared to previous ansatz\cite{Wahl:2017} even with block size $l=6$. Empirically, we observe that the running time of the binary uTNC is almost similar to regular uTNC with $\tau=2$ and $l=4$. The better performance of binary ansatz is understood from having deeper layers of unitary gates, as mimicking more accurately localization length than a simple regular structure.

\begin{table}[tp]
\caption{ Comparison of the cost functions for the tensor-network structures shown in Fig.~\ref{fig:MERAarchitecture}(c-f). The parameter $\bar{D}$ stands for the PEPS bond dimension. The energy variance is averaged over 20 realizations.
}
\begin{ruledtabular}
\begin{tabular}{lrrrrr}
$architecture$&$N$& $\sigma^2/2^N$&$W$& \\
\colrule
$regular$  & $(16,32)$ & $(0.1,0.4)$&$8$ \\
$iregular$  & $(16,32)$ & $(0.1,0.4)$&$8$   \\
$binary$ & $(16,32)$ & $(0.05,0.08)$ &$8$ \\
$ternary$  & $(16,32)$ & $(0.03,0.07)$ &$8$ \\
\colrule
\colrule

$architecture$&$l_y$& $\parallel M-$QR$ \parallel$&$\bar{D}$& \\
\colrule

$regular$ & $10$ & $1\times 10^{-2}$ &$3$\\
$irregular$  & $10$ & $1\times 10^{-2}$  &$3$ \\
$binary$  & $10$ & $4\times 10^{-4}$ &$3$ \\
$ternary$  & $10$ & $2\times 10^{-4}$ &$3$ \\
\end{tabular}
\end{ruledtabular}
\label{table1}
\end{table}

We now compare the tensor-network structures, shown in Fig.~\ref{fig:MERAarchitecture}(c-f), by studying the energy variance and the tensor-network $QR$-decomposition. In Tab.~\ref{table1}, we have presented the results for disorder strength $W=8$ and the PEPS columns with length $l_y=10$ and bond dimension $\bar{D}=3$. The ternary uTNC provides the best accuracy as it includes more variational parameters (with an entangled structure) compared to other ones. The iteration number to obtain converged results is related to variational parameters, thus the running time to perform the ternary uTNC is a factor of $\sim 2$ larger than the binary uTNC, while it remains (almost) the same for the binary, regular and irregular uTNC. Surprisingly, the irregular and regular uTNC provides the same level of accuracy, as one might expect to see the irregular structure (due to a non-local connection of tensors) result in better performance. A global energy optimization might change these results (compared to the local optimization technique used here) as the irregular structure includes non-local connections.

We end this section by presenting some simple benchmark results for time evolution. We study the time-evolution of the entanglement entropy after a local quench. We choose the initial state to be a product state in $s_z$-basis with a spin-flip operator on the $i$th site, i.e., $s^{i}_x  |0\rangle^{\otimes N}$. We use the methods developed in Sec.\ref{Sec:timeevaluation} to study the dynamics of  entanglement entropy: we plot the time-dependent entanglement entropy $S(t)$, obtained from the von Neumann entropy of $e^{-i t H} |\psi\rangle$, as a function of time $t$ in Fig.~\ref{fig:Conv}(c, d). The result is presented for the Heisenberg chain with random magnetic fields with $W=5$ (one realization). We observe that by increasing the accuracy of the uTNC ansatz, we can reproduce the exact results. A relative error of order $\sigma^{2}/2^N\sim 10^{-3}$ is enough to qualitatively mimic the exact time-evaluation result. 


\section{Conclusion} 
In this paper, we have studied the optimization methods of uTNC with different network structures. We discussed in detail an efficient local optimization of energy variance based on MERA techniques. We studied a CG method with an effective line search and empirically show that it provides the most efficient scheme compared to the linearizing algorithm and the Armijo-based algorithm. The method makes the ansatz faster by a large prefactor that linearly increases with the system size and circuit depth. Empirically, it is noted that the uTNCs with a MERA-like structure provide the best performance, compared to previously proposed structures, as observed by the benchmarking results presented for a tensor-network $QR$-decomposition and estimating the eigenspectra of a system exhibiting MBL. We also presented a time-evolution algorithm based on uTNC to evaluate the time-dependent entanglement entropy of a system exhibiting MBL after a local quench, as studied for the Heisenberg model in a disordered magnetic field.  

Future research may focus on developing global optimization schemes for (non-local) network structures (especially in two dimensions) and its applications in the canonical PEPS ansatz, studying phase transition in MBL systems and real-time evolution. The network structures and optimization schemes introduced in this paper could be used in improving quantum circuit algorithms on the noisy intermediate-scale quantum (NISQ) computers.\cite{McArdle:2019, Motta:2020, lin2020real}

\acknowledgments
This work was supported by the US Department of Energy, Office of Science, via award no
DE-SC0019374. We have used library Uni10~\cite{Kao:2015} to perform tensor-network ansatz. We thank F. Pollmann for the initial ideas of this work and also thank D. N. Sheng, A. Langari, A. T. Rezakhani and G. K. Chan for helpful discussions. We appreciate F. Pollmann and P. Helms for their useful comments and for reading the manuscript.
        
\bibliography{references} 
 
\end{document}